# Impact of Anisotropic Exchange on M-H Loops: Application to ECC Media

M.L. Plumer[1], M.C.Rogers[1], and E. Meloche[2]

[1]Department of Physics and Physical Oceanography, Memorial University of Newfoundland, St. John's, NL A1B 3X7, Canada
[2]Seagate Technology, 7801 Computer Ave. S., Bloomington, MN 55435, USA

**Micromagnetic simulation results on Co-based recording media are presented which examine the impact of a modified near-neighbor exchange interaction between grains of the form $J^z(M_i^z M_j^z)$, reflecting the hexagonal crystal symmetry. Both out-of-plane and in-plane M-H loops are calculated, with an emphasis on a model fit to data reported by Wang *et al.* [*IEEE Trans. Magn.* vol. 43, 682 (2007)] on exchange coupled composite perpendicular media. The principle effect of $J^z$ is to increase the coercivity and slope of both hard and soft layers. Improved agreement with experimental data for the in-plane loops is achieved by assuming a substantial value for $J^z$. Possibilities for measurement of $J^z$ through spin-wave excitations are discussed. Thermal fluctuation effects are also examined through simulations of the magnetization vs temperature.**

## I. Introduction

The amorphous nature of sputtered cobalt-based granular thin films for recording media offers a number of challenges to model key features relevant to increasing areal density and thermal stability[1]. The fundamental ingredient to modern micromagnetics is the energy expressed as a functional of the magnetic moment vectors of interacting grains, E[$\mathbf{M}_i$]. An important aspect of the energy is that it should reflect the symmetry of the granular crystallites, typically hcp for Co-based films. It is important to recall that the magnetism on a granular scale, which forms the basis for micromagentics, has its origins in atomic-level crystal field and exchange effects. The uniaxial anisotropy associated with hexagonal crystal systems gives rise to the well-known single-site contribution to the energy of the form $-K(M_i^z)^2$. For $K$>0, this term is minimized by $\mathbf{M}||\mathbf{c}||\mathbf{z}$, giving rise to the desired perpendicular anisotropy for crystallites having the **c** axis perpendicular to the film plane. However, hexagonal symmetry also allows for inter-site anisotropy terms of the form $-J^z(M_i^z M_j^z)$. Such *anisotropic exchange* has been examined in the cases Co and Fe ions [2,3] and has previously been included in a micromagnetic calculation of domain structures in MnAs films [4].

The present work reports on an initial investigation of such anisotropic exchange in a micromagnetic model relevant to modern Co-based perpendicular recording media. Emphasis is placed on the impact of this interaction on the relative shapes of easy and hard-axis M-H loops. A specific application is made to the exchange coupled composite (ECC) media reported by Wang *et al.* [5]. The results of fitting the model to experimental M-H loops for both hard and soft layers individually are shown. Comparisons with data on loops in the case of exchange coupled films are also made. The effect of anisotropic exchange on thermal stability is discussed and some conclusions are also made regarding the benefits of ECC media.

## II. Micromagnetic Simulations

Simulations were performed using the commercial Landau-Lifshitz-Gilbert (LLG) micromagnetic simulator [6] with rectangular-prism grains on a square lattice with 32x32 grains and periodic boundary conditions. Both domain-wall (incoherent) and rotational (coherent) reversal behavior can occur within this discretization scheme. Anisotropic exchange is incorporated as a modification of the usual isotropic exchange term by writing

$$E_{ex} = -J\sum_{ij}(M_i^x M_j^x + M_i^y M_j^y + mM_i^z M_j^z) \qquad (1)$$

where the sum is over near-neighbor grains, m>1 induces uniaxial anisotropy and m<1 creates planar anisotropy. Note that anisotropy of this type does not occur in crystals with cubic symmetry. For the exchange coupling between hard and soft layers in our model of ECC media, it was always assumed that m=1. The usual exchange parameter is expressed as $A=Jd^2$, where $d^2$ is the cross-sectional area of a grain. Single-ion uniaxial anisotropy ($K$), fourth-order uniaxial anisotropy ($K_2$) [7], and magnetostatic interactions are also included. Zero temperature calculations were performed using a Suzuki-Trotter rotation-matrix method [8] with a damping parameter α=1. Finite temperature effects were included through the usual Langevin stochastic term [9] with an Euler integration routine, α=0.2, and a time step of 0.1 ps. Various quantities such as the saturation magnetization $M_s$, anisotropy magnitude ($K$) and direction (κ) were assumed to have a Gaussian distribution characterized by a given standard deviation, SD. We note that distributions in magnetic properties also serve to account for distributions in geometrical features, such as grain size (not explicitly included in the present model).

## III. Impact of Anisotropic Exchange on M-H Loops

In order to explore the impact of anisotropic exchange on M-H loops, a generic model of single-layer perpendicular media was assumed. Grains with dimensions 7x7x20 nm$^3$ were used, having the following properties: $M_s$=4.5x10$^5$ A/m (SD=10%), $A$=0.05 – 0.15 x10$^{-11}$ J/m, anisotropic exchange parameter m=1 or 2, $K$=1.0x10$^5$ J/m$^3$ (SD=10%), as well as $\kappa_{SD}$=4$^0$. Figs. 1 and 2 illustrate some key features, with the applied field H normalized by the anisotropy field $H_K$=2K/$M_s$. With m=2, both the coercivity and slope increase for out-of-plane (easy axis) loops compared to the isotropic case, m=1. Reducing the exchange parameter $A$ has the effect to reduce the coercivity ($H_c$) and the slope, as expected [10]. In the case of in-plane loops, the slope decreases substantially for m>1 with little effect on the coercivity. This is a noteworthy result. These features are consistent with the expectation that increasing m will serve to both increase the effective anisotropy as well as the exchange interaction if $M_z \neq 0$.

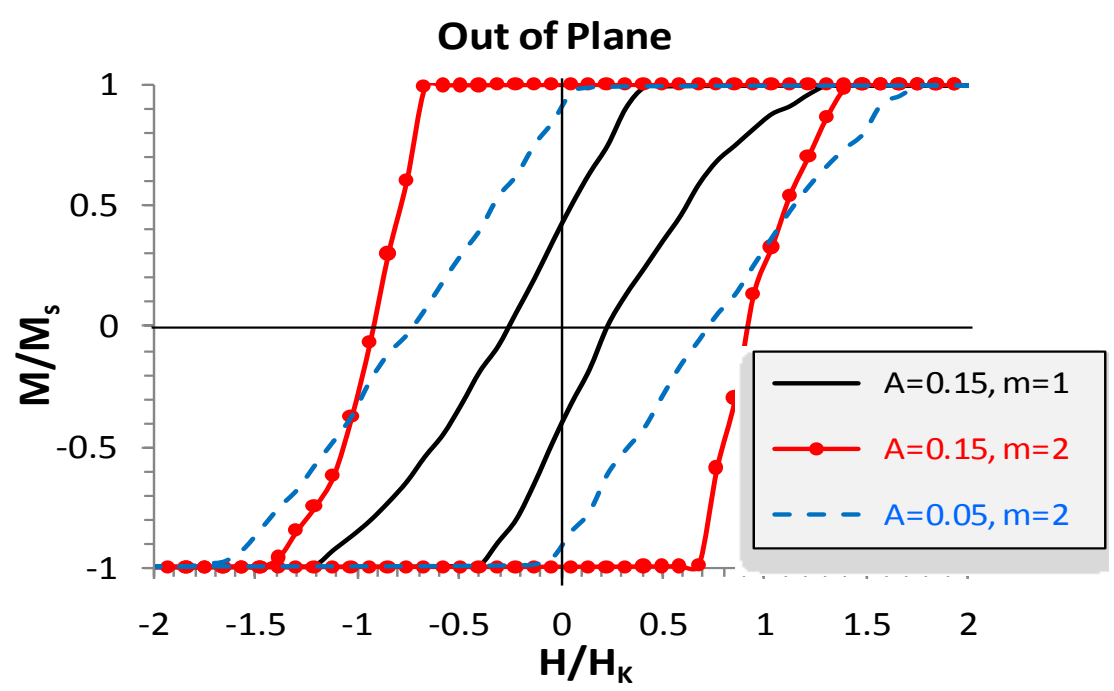

Fig. 1. Easy-axis M-H loop illustrating some effects of anisotropic exchange in perpendicular media. The overall exchange parameter $A$ is in units of 10$^{-11}$ J/m.

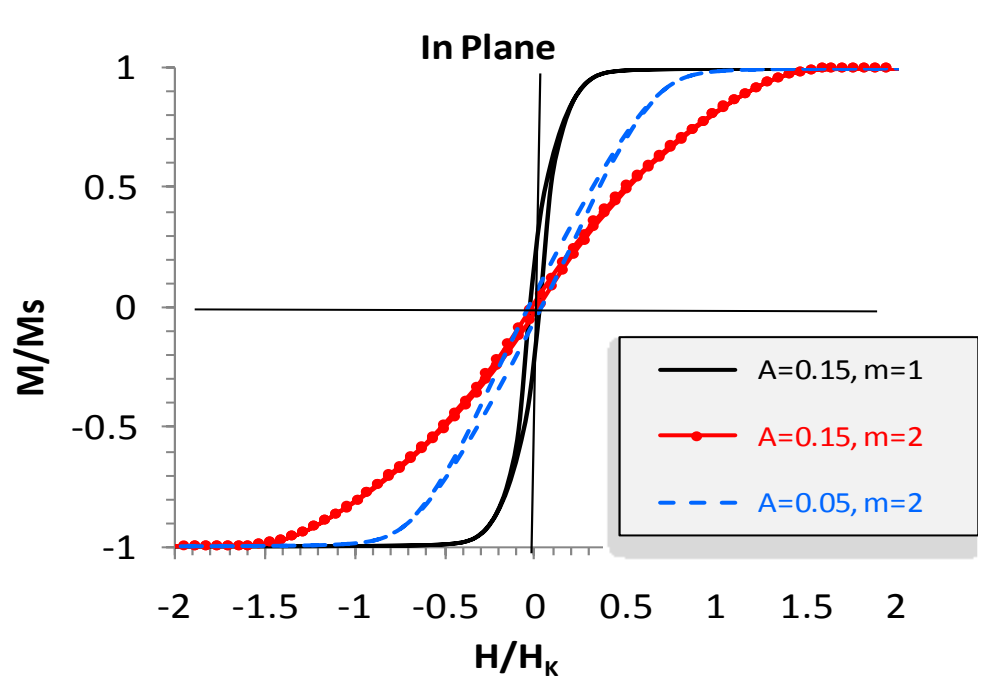

Fig. 2. As in Fig. 1 but with the applied field in the plane.

## IV. MODEL OF THE ECC MEDIA OF WANG *ET AL*.

### *A. Hard Layer*

Fig. 3 shows the results of fitting model parameters for both cases with isotropic and very anisotropic (m=5), exchange in an attempt to reproduce the out-of-plane M-H loop data for the hard layer of Wang et al. [5]. Grains with dimensions 7x7x20 nm$^3$ were again used along with $M_s$=4.5x10$^5$ A/m (SD=10%). For the case with m=5, fourth-order uniaxial anisotropy ($K_2$=$K$/4) was also included with the same SD as assigned to $K$. A large easy-axis-direction SD of $\kappa_{SD}$=10$^0$ was required (for both m=1 and m=5) to reproduce the rounding observed near the nucleation field ($H_n$). Larger $\kappa_{SD}$ also has the effect of decreasing the easy-axis $H_c$ and increasing the hard-axis $H_c$. This value of $\kappa_{SD}$ required to reproduce the data is much larger than the reported $\Delta\theta_{50}$ = 3$^0$ - 4$^0$. The reasons for this difference are not entirely clear but may be related to grains having a core-shell structure, with a large distribution in the shell anisotropy axes [11]. In the case of m=1, the fitted values of exchange and anisotropy were found to be $A$=0.15x10$^{-11}$ J/m and $K$=2.0x10$^5$ J/m$^3$ (SD=20%). For the m=5 case, the values $A$=0.02x10$^{-11}$ J/m and $K$=1.45x10$^5$ J/m$^3$ (SD=20%) gave the best fit to the data.

Using these fitted parameters, somewhat better agreement with the in-plane experimental data is found in the case of m=5, as shown in Fig. 4. This can be understood since the exchange parameter $A$ is reduced from its m=1 fitted value, thus giving a smaller slope as seen experimentally. However, the overall agreement is significantly worse than in the easy-axis case of Fig. 3. Note the large experimental $H_c$, not expected from coherent rotational switching usually assumed for hard-axis loops, as well as the lack of saturation in M.

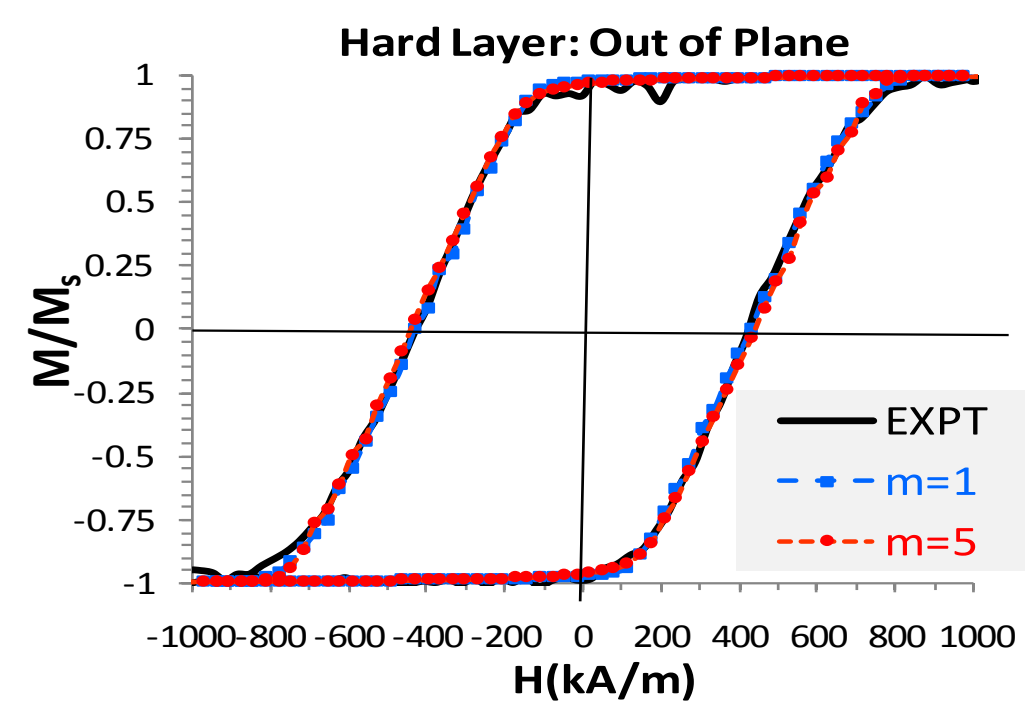

Fig. 3. Model fit to the out-of-plane hard layer M-H loop data of Wang *et al*. [5] for both isotropic and anisotropic exchange.

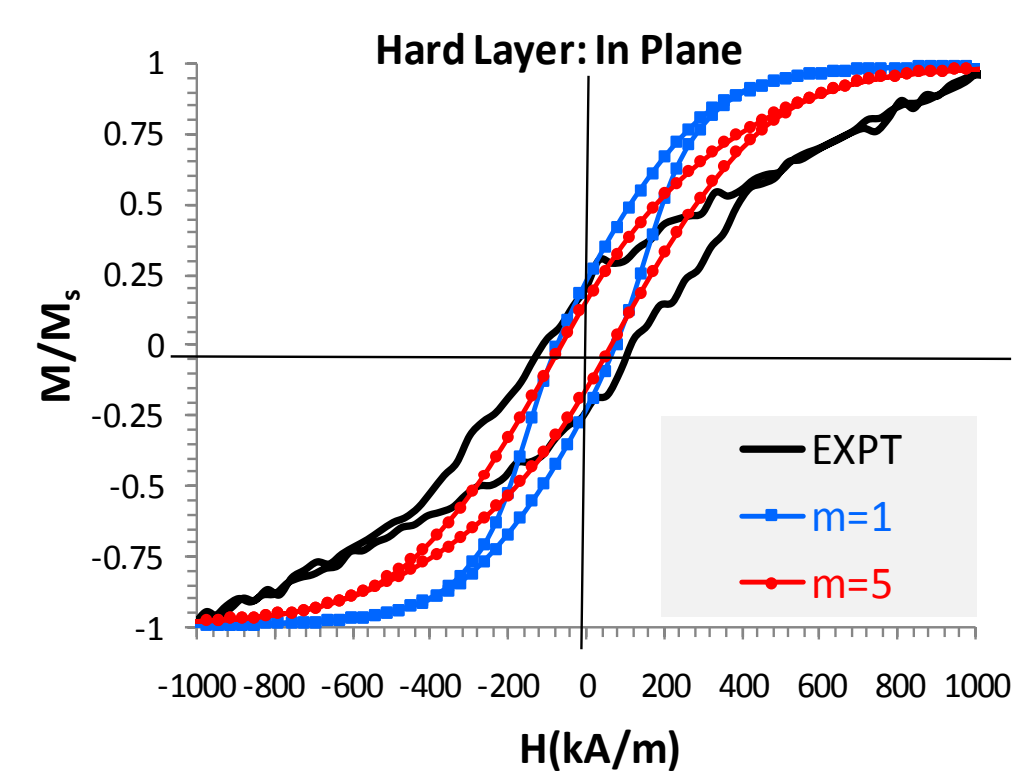

Fig. 4. In-plane hard layer M-H loops resulting from using out-of-plane fitted parameters from Fig. 3. Experimental data from Wang *et al.* [5] is also shown.

### *B. Soft Layer*

A reduction in film thickness as well as Pt content (relative to the hard layer) serves to increase the amount of stacking faults [5]. The effect is to reduce $H_c$ (by about a factor of 10) in the soft layer and produce a more isotropic film, with a larger anisotropy axis distribution. Model parameters were adjusted to best reproduce the experimental soft layer out-of-plane M-H loop [5]. Here, the layer thickness was reduced to 8 nm and a saturation magnetization value $M_s$=8x10$^5$ (SD=10%) was assigned. For the case m=1, the values $\kappa_{SD}$=17$^0$, $A$=0.48x10$^{-11}$ J/m, K=3.8x10$^5$ J/m$^3$ (SD=10%) fit the data well, as shown in Fig. 5. Assuming m=5, the fitting procedure yielded values $\kappa_{SD}$=20$^0$, $A$=0.075x10$^{-11}$ J/m, $K$=1.25x10$^5$ J/m$^3$ (SD=20%) which notably required larger distributions, which are known to degrade playback signal quality [12]. Note that the anisotropy fields, $H_K$=2K/$M_s$, are about the same for hard and soft films but the intrinsic anisotropy parameter D=K/$M_s^2$ is smaller in the soft film by about a factor of two. The reduction in soft-layer Hc is mainly due to a larger $M_s$ and $\kappa_{SD}$.

Using these parameter values, modeled in-plane loops, compared with the experimental data, are shown in Fig. 6. As with the hard layer, agreement with the data is much worse but again the model with anisotropic exchange appears somewhat better. Note that the experimental values of $H_c$ in both Figs. 5 and 6 are similar, suggesting that a combination of domain-wall and rotational reversal occurs for both hard and easy-axis loops. The model again appears to underestimate the anisotropy magnitude (see Conclusions).

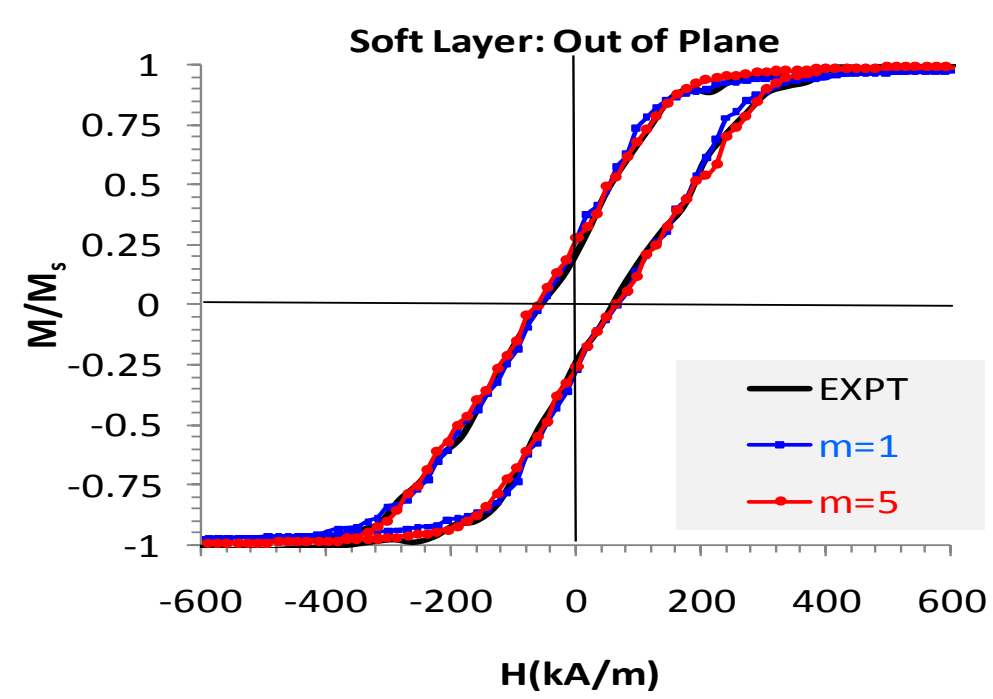

Fig. 5. Model fit to the out-of-plane soft layer M-H loop data of Wang *et al*. [5] for both isotropic and anisotropic exchange.

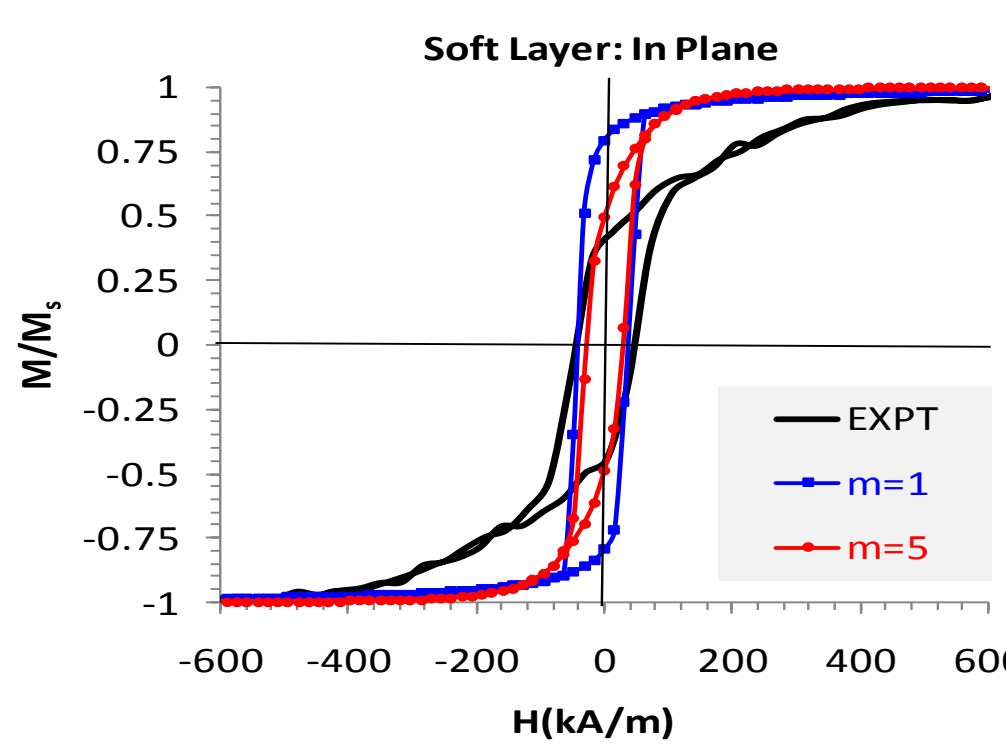

Fig. 6. In-plane soft layer M-H loops resulting from using out-of-plane fitted parameters from Fig. 3.

## C. Dual Layer

ECC media is fabricated by controlled exchange coupling across soft and hard films through varying the thickness $t_s$ of a spacer layer, composed of Pt in the work of Ref. [5]. Parameters identified by the above analysis were used to model such coupled media. An inter-layer exchange parameter, $A_i$=0.0 – 1.5x10$^{-11}$ J/m was added (with [13] m=1) and varied in an attempt to mimic the effects of the experimentally controlled $t_s$=0.7 – 3.6 nm.

Some example easy-axis loops are shown in Figs. 7 and 8 for the cases of m=1 and m=5, respectively, where $A_i$ =0 and $A_i$ = 1.0x10$^{-11}$ J/m were used. Data from Ref. [5] for the thinnest and thickest spacer layers are also shown. The general trends that $H_c$ is little affected, but that slopes decrease (increasing saturations fields $H_s$) with increasing $t_s$, is reproduced by the model using a smaller $A_i$. It is also evident that the model with anisotropic exchange again gives results in better agreement with the data. Overall, the observed effect of reduction in $H_c$ by about 40% from the single-layer value is reproduced by the model. Hard-axis loops were also calculated (not shown) which exhibit coercive fields of about 70 kA/m for the m=1 case and 25 kA/m for the m=5 case.

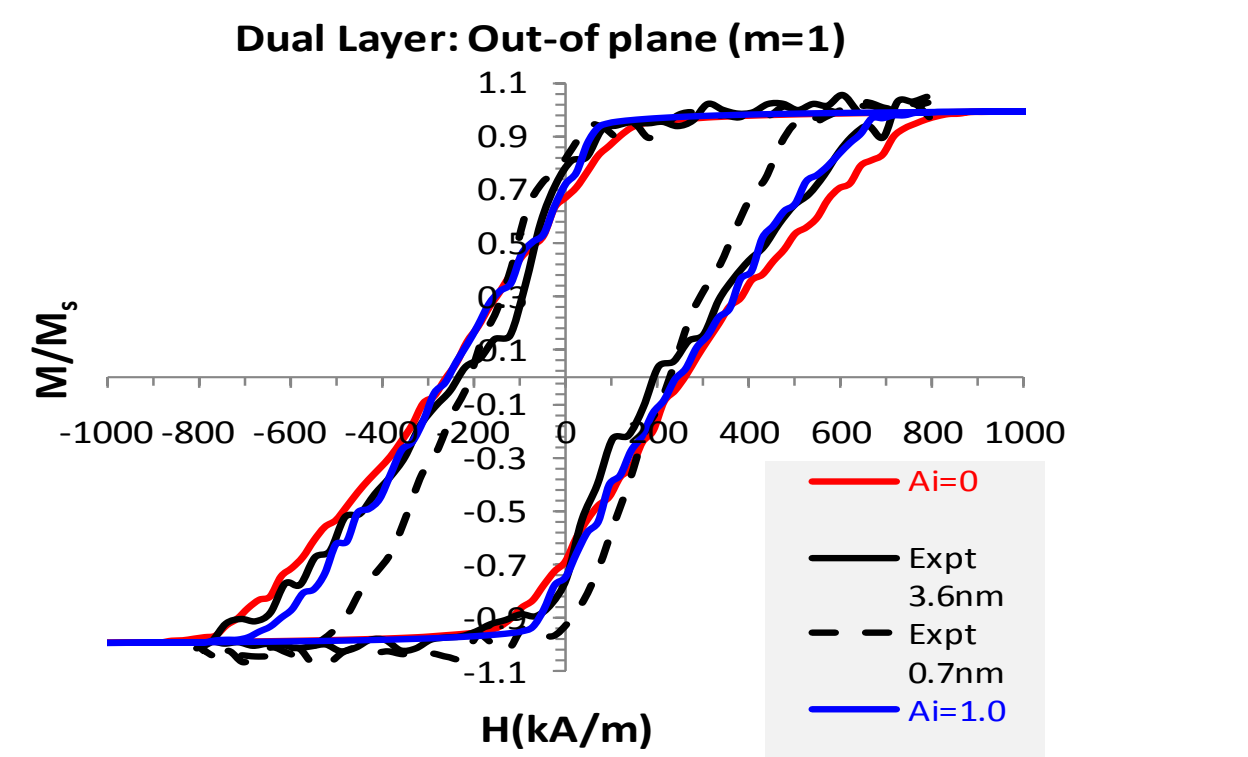

Fig. 7. Easy-axis dual-layer media loops comparing m=1 modeled and experimental results [5]. Interlayer exchange $A_i$ is in units of 10$^{-11}$ J/m.

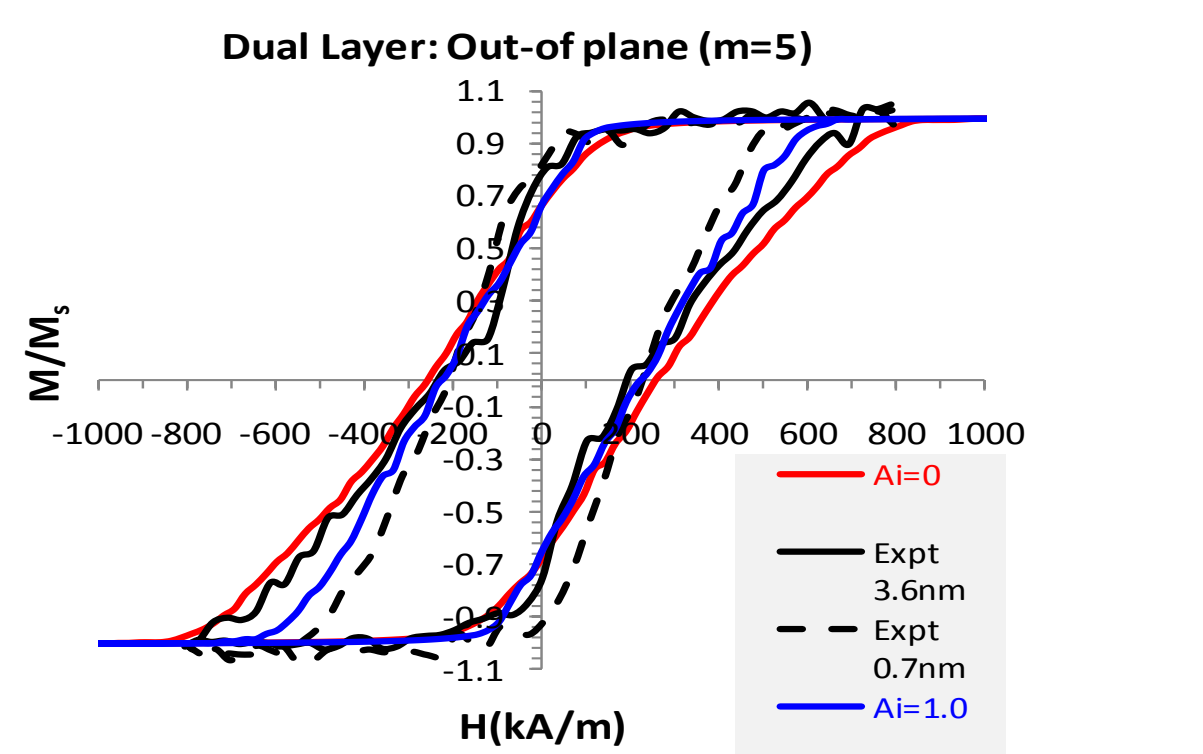

Fig. 8. As in Fig. 7 but with m=5 modeled results.

## V. MAGNETIZATION VS TEMPERATURE.

A useful indication of the impact of thermal fluctuations in magnetic systems is the reduction in M with increasing temperature [14]. Examination of simulation results for M vs time exhibit an equilibrium value. For the finite-size systems (as studied here), the transition to paramagnetism, $T_c$, is indicated by a large reduction in M from its zero-temperature value (reaching zero only for infinite systems) [15].

Finite-temperature simulations were performed on the single and dual layers using model parameters discussed above to obtain the magnitude of **M** vs T. That thermal effects were properly treated in the model was verified by reproducing the results of Ref. [14] as well as corresponding MC simulations. Simulation times of 150 ns were found to be adequate for the purpose of estimating M(T). We estimate (somewhat arbitrarily) $T_c$ by the relation $M(T_c) = 0.15M(0)$.

Results for the hard-layer magnetization, in single-layer cases m=1and m=5, as well as the case of m=1 for dual layers with $A_i$=0.1 and 1.0x10$^{-11}$ J/m, are shown in Fig. 9. In the single-layer cases, $T_c$ for both is estimated to be approximately 1700 K. Interlayer exchange has little impact on $T_c$ when the hard layer is coupled to the soft layer. It is estimated to be about 1200 K for both dual-layer cases studied. A more accurate estimate of $T_c$ will require simulations using larger lattices.

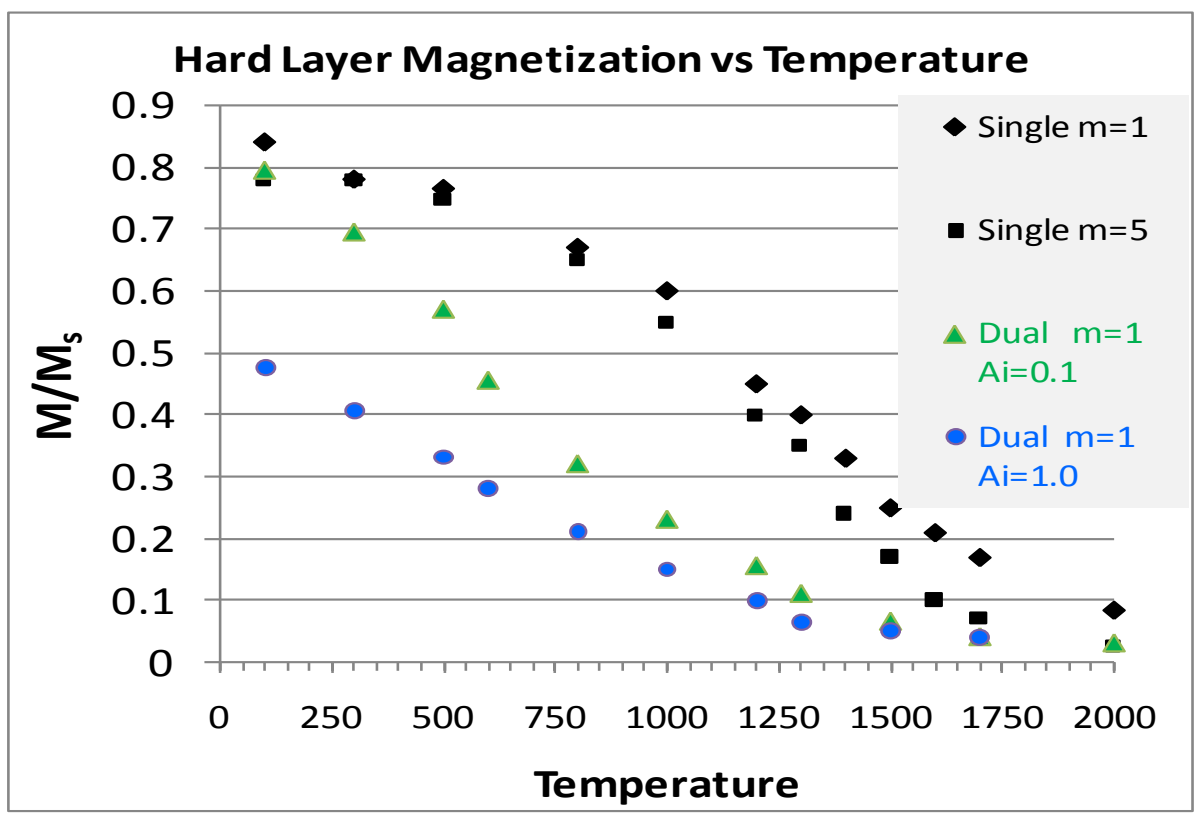

Fig. 9. Hard layer magnetization as a function of temperature for single and dual layer films. Interlayer exchange $A_i$ is in units of $10^{-11}$ J/m.

## VI. Spin Waves

If magnetic interactions in sputtered cobalt-based films do include significant exchange anisotropy, it is not immediately obvious how this can be measured. One possibility is through spin-wave dispersion curves as a function of wave vector **q**. Single-ion anisotropy *K* appears as a gap at q=0 and the curve shape determines the exchange constant *J*. A simple calculation based on the torque equation (LLG equation with α=0) shows that for perpendicularly magnetized thin films, *K* and $\mathcal{J}$ simply add together to increase the gap [16]. However, if a strong magnetic field H is applied in the film plane, anisotropic exchange also affects the shape of the dispersion as in the following relations (for a square lattice with inter-atomic separation distance *a*):

$$\omega_q^2 = [H + 4MJ(1 + f_q)][H - 2KM + 4MJ\left(1 - mf_q\right)] \quad (2)$$

$$f_q = (cosq_x a + cosq_y a)/2 \quad (3)$$

The anisotropy parameter m could in principle be used as a distinguishable fitting parameter along with *J* and *K*.

## VII. Discussion and Conclusion

This work has shown that anisotropic exchange may play a role in micromagnetic models of perpendicular media, especially when faced with the challenging task of reproducing both easy *and* hard axis M-H loops. We note that an approach based on mimicking long experimental sweep-rate effects by assuming an elevated temperature [17] yields larger fitted anisotropy values and better agreement for in-plane loops (to be reported elsewhere).

Some conclusions from the present work may be made regarding the relative benefit of ECC media. Although there is a reduction in $H_c$, we have found a substantial reduction in $T_c$ as well when the hard layer is coupled to a soft layer (indicating larger thermal fluctuations). Taking the ratio of $H_c/T_c$ as a figure of merit, our results yield approximately 3.0 kA/(mK) for the single-layer and 2.5 for the dual-layer cases, where a smaller number is advantageous for a combination of writability and thermal stability. The modeled ECC media is seen to yield about a 20% advantage by this criterion.

## Acknowledgment

We thank W. Shen and J.-P. Wang for providing detailed data associated with their M-H loops [5] and M. Scheinfein for numerous correspondence and modifications to his code [6]. This work was supported by NSERC of Canada.

## References

[1] *The Physics of Ultra-High-Density Magnetic Recording*, Eds. M. Plumer, J. van Ek and D. Weller, Springer-Verlag, 2001.
[2] C.A. Bates *et al*., "Exchange interactions between two cobalt ions and a calculation of the g-shift," *J. Phys. C,* vol. 9, p. 1511, 1976.
[3] R. Skomski, A. Kashyap, J. Zhou, and D.J. Sellmeyer, "Anisotropic exchange," *J. Appl. Phys*., vol. 97, p.10B302, 2005.
[4] D.M. Schaadt, R. Engel-Herbert, and T. Hesjedal, "Effects of anisotropic exchange on the micromagnetic domain structures," *Phys. Stat. Sol. B*, vol. 244, p. 1271, 2007.
[5] J.-P. Wang, W. Shen, and S.-Y. Hong, "Fabrication and characterization of exchange coupled composite media," *IEEE Trans. Magn*., vol. 43, p. 682, 2007.
[6] http://llgmicro.home.mindspring.com/
[7] L. Guan, Y.-S. tang, B. Hu, and J.-G. Zhu, "Thermal stability enhancement of perpendicular media with higher-order uniaxial anisotropy," *IEEE Trans. Magn.*, vol. 40, p. 2579, 2004.
[8] S.-H. Tsui and D.P. Landau, "Spin Dynamics," *Computer Simulations*, Jan./Feb. p. 72, 2008.
[9] W.F. Brown, "Thermal Fluctuations of a Single-Domain Particle," *Phys. Rev*., vol. 130, p. 1677, 1963.
[10] A. Nakamura, M. Igarashi, M. Hara, and Y. Sugita, "M-H loop slope and recording properties of perpendicular media," *Jap. J. Appl. Phys*., vol. 43, p. 6052, 2004.
[11] M.P. Wismayer *et al*., "Using small-angle neutron scattering to probe the local magnetic structure of perpendicular magnetic recording media," *J. Appl. Phys*., vol. 99, p. 08E707, 2006.
[12] M.L. Plumer and J. van Ek, "Perpendicular recording model of medium and head field saturation effects," *IEEE Trans. Magn.*, vol. 38, p. 2057, 2002.
[13] The value m=1 is used here for simplicity. Other cases will be explored in future work.
[14] X. Feng and P.B. Visscher, "Course-graining Landau-Lifshitz damping," J. *Appl. Phys*., vol. 89, p. 6988, 2001.
[15] V. Privman, editor, *Finite size scaling and numerical simulation of statistical systems*, World Scientific, Singapore, 1990.
[16] T. Nguyen, private communication.
[17] J. Xue and R.H. Victora, "Micromagnetic predictions for thermally assisted reversal over long time scales," *Appl. Phys. Lett*., vol. 77, p. 3432, 2000.